\documentclass[twocolumn,showpacs,preprintnumbers,amsmath,amssymb,superscriptaddress, aps, prb, floatfix]{revtex4-1}
\pdfoutput=1
\usepackage{graphicx}
\usepackage{xcolor}
\usepackage{hyperref}
\usepackage{bm}

\def\Im{\mathop{\rm Im}}
\def\Tr{\mathop{\rm Tr}}

\begin{document}

\title{Electronic structure and magnetism of samarium and neodymium adatoms on free-standing
  graphene}

\author{Agnieszka L. Kozub} \email{kozub@fzu.cz}
\affiliation{Institute of Physics, Czech Academy of Sciences, Na Slovance 2, 182 21 Prague, Czech Republic}
\affiliation{Faculty of Applied Physics and Mathematics, Gdansk University of
Technology, Narutowicza~11/12, 80-233 Gdansk, Poland}
\author{Alexander B. Shick}
\affiliation{Institute of Physics, Czech Academy of Sciences, Na Slovance 2, 182 21 Prague, Czech Republic}
\author{Franti\v sek  M\' aca}
\affiliation{Institute of Physics, Czech Academy of Sciences, Na Slovance 2, 182 21 Prague, Czech Republic}
\author{Jind\v{r}ich Koloren\v{c}}
\affiliation{Institute of Physics, Czech Academy of Sciences, Na Slovance 2, 182 21 Prague, Czech Republic}
\author{Alexander~I.~Lichtenstein}
\affiliation{Institute of Theoretical Physics, University of Hamburg, Jungiusstra\ss e 9, 20355 Hamburg, Germany}

\begin{abstract}
The electronic structure of selected rare-earth atoms adsorbed on a free-standing graphene was investigated using methods beyond the conventional  density
functional theory (DFT+U, DFT+HIA and DFT+ED). The influence
of the electron correlations and the spin-orbit coupling on the
magnetic properties has been examined.
The DFT+U method predicts both atoms to carry
local magnetic moments (spin and orbital) contrary to 
a nonmagnetic $f^6$ ($J=0$) ground-state
configuration of Sm in the gas phase.
 Application of DFT${}+{}$Hubbard-I (HIA)
and DFT${}+{}$exact diagonalization (ED) methods cures this problem, and yields a
nonmagnetic ground state with six $f$ electrons and $J=0$
for the Sm adatom. Our calculations show that Nd adatom remains magnetic, with four 
localized $f$ electrons and $J=4.0$. These conclusions could be verified by STM and XAS experiments.

\end{abstract}
\date{\today}
\maketitle

\section{Introduction}

Adsorption of  atoms and molecules provides a way to control and modify
the electronic properties of graphene~\cite{Misha2012}. Adsorption of
alkali and transition metals on graphene was investigated extensively in
recent years.\cite{eelbo2013,sessi2014,wehling2011,virgus2014}
 There are much less studies of interaction between
rare-earth atoms and graphene. 
Since the bonding character of the $sp$ elements and transition metals is
different from  that of strongly localized $4f$ metals, a different
behavior of the
rare-earth atoms adsorbed on graphene is expected. In the pioneering
work~\cite{Liu2010}, the first-principles theory has been applied to
several rare-earth adatoms on graphene, together with the scanning
tunneling microscopy (STM) experiments.
It was shown that the hollow site of graphene is the energetically
favorable adsorption site for all the rare-earth adatoms. Magnetic moments
have been reported for all adatoms studied.

Accurate description of the electronic and magnetic properties of the $f$-electron
systems remains a challenge in condensed matter physics. The standard
density-functional theory (DFT)
proves to be inadequate due to the self-interaction error~\cite{Cohen2008}. For this reason,
theories like self-interaction correction~\cite{Perdew1981}, hybrid functionals~\cite{Becke1993} 
or treatment of the 4$f$-shell as  core-like~\cite{Skriver1985} have been explored. In Ref.~\onlinecite{Liu2010},
the $f$-states of the rare-earth adatoms were treated
as a part of the atomic core and were fixed in a given configuration. That places some limits on 
the validity of acquired conclusions about the magnetic character of the $f$-manifold.

In this paper, we re-examine the electronic and magnetic structure of
two rare-earth adatoms (Sm and Nd) on graphene 
making use of the rotationally invariant formulation of the DFT+U
method~\cite{LAZ1995}. In order to incorporate the  dynamical electron
correlations, we employ the exact diagonalization (ED) method to solve
a multi-orbital single-impurity Anderson model~\cite{Hewson} whose parameters are 
extracted from DFT calculations. This method is conceptually similar
to earlier calculations of bulk rare-earth materials.\cite{thunstrom2009,Locht2016}

In Sec.~\ref{sec:methods} we describe the DFT+U and DFT+ED methods 
which we use to calculate the electronic structure and magnetic properties of the adatoms on graphene. 
Special attention is paid to modifications of the DFT+U due to the
spin-orbit coupling (SOC). In Sec.~\ref{sec:Sm} we describe
the results of the DFT+U, DFT+Hubbard I (HIA) and DFT+ED calculations of Sm adatom on graphene (Sm@GR). 
It is shown that the $f^6$ shell of Sm with the nonmagnetic singlet ground state cannot be 
described correctly by DFT+U.  The use of DFT+HIA and DFT+ED solves this problem.
In Sec.~\ref{sec:Nd} we address the electronic and magnetic character of a rare-earth adatom with
the local moment, taking as an example Nd adatom on graphene (Nd@GR).
A comparison between DFT+U and DFT+HIA is given. Reasonable agreement between the DFT+U
and DFT+HIA  $f$-projected density of states (DOS) is demonstrated.

\section{Computational methods}
\label{sec:methods}
The conventional band theory fails to correctly describe the strongly
localized 4$f$ states due to
the oversimplified treatment of electron correlations, as it
is often seen in the applications of DFT to $f$-electron materials.
Here, we use the correlated band theory (DFT+U) method, which
consists of DFT augmented by a correcting energy of a multiband
Hubbard type.

\begin{figure}
\centerline{
\includegraphics[angle=0,width=0.5\columnwidth,trim=0 60 0 40,clip]{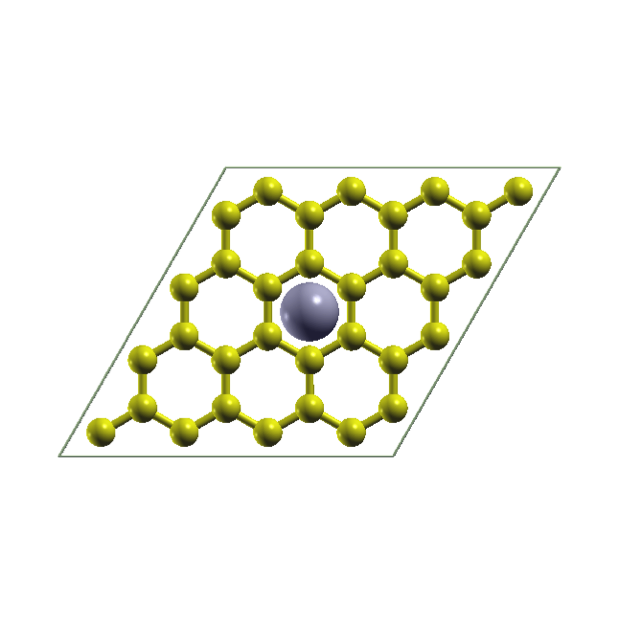}
\includegraphics[angle=0,width=0.5\columnwidth,trim=0 60 0 40,clip]{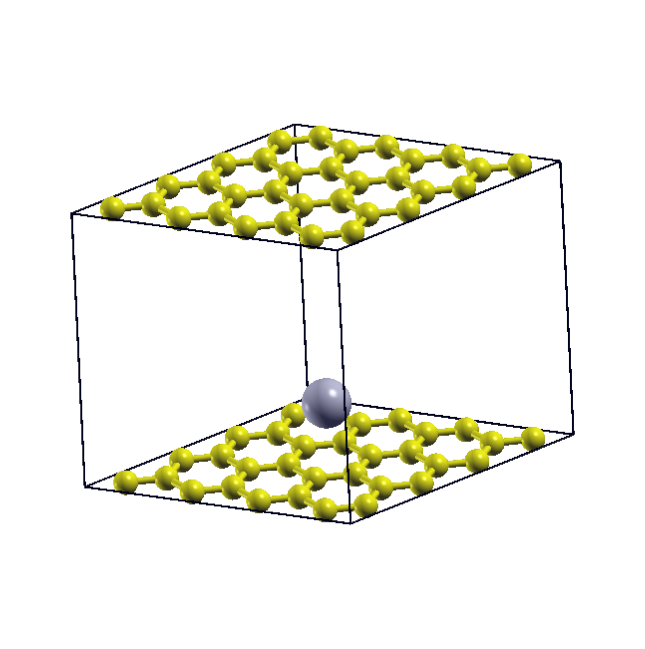}}
\caption{Schematic supercell model for rare-earth impurity on graphene.}
 \label{fig1}
\end{figure}

In order to describe the structural, electronic and magnetic
properties of the rare-earth adatoms
on graphene, we use the supercell shown in Fig.~\ref{fig1}. This
$4\times4\times1$ supercell includes 32 carbon atoms,
and the rare-earth adatom is placed in the hexagonal hollow position.
First, the structure relaxation was performed employing the standard
Vienna \emph{ab initio} simulation package (VASP)~\cite{Kresse1996} together with
the projector augmented-wave method (PAW)~\cite{Blochl1994}  without
SOC. We used the DFT+U method with the exchange-correlation functional
of Perdew, Burke and Ernzerhof (PBE).\cite{PBE1996} The Coulomb~$U$ values of 6.76 eV (Nd) and
6.87 eV (Sm), and the exchange~$J$ of 0.76 eV were used, which are in
the commonly accepted range of $U$ and $J$ for the
rare earths\cite{vdMarel1988}. The optimal heights for the rare-earth adatoms above the
graphene sheet are found as $h_{\rm Sm}=4.58$ bohr and $h_{\rm
  Nd}=4.55$ bohr.

The structural information obtained from the VASP simulations was used
as an input for further electronic-structure calculations that employ
the relativistic version of the full-potential linearized augmented
plane-wave method (FLAPW)~\cite{wimmer81}, in which the SOC is included in a self-consistent second-variational
procedure~\cite{shick1997}. This two-step approach
synergetically combines the speed and efficiency of the highly optimized
VASP package with the state-of-the-art accuracy of the FLAPW method.

\subsection{DFT+U with spin-orbit coupling}

When the spin-orbit coupling is taken into account, the
spin is no longer a good quantum number, and the
electron-electron interaction energy $E^{\rm ee}$ in the DFT+U
rotationally-invariant total-energy functional \cite{LAZ1995} has to
be modified~\cite{solovyev1998} to
\begin{equation}
\label{eq:eeHam} E^{\rm ee} = \frac{1}{2}  \sum_{\gamma_1 \gamma_2
\gamma_3 \gamma_4} n_{\gamma_1 \gamma_2} \Big( V^{\rm ee}_{
\gamma_1 \gamma_3; \gamma_2 \gamma_4} - V^{\rm ee}_{\gamma_1
\gamma_3;\gamma_4 \gamma_2} \Big) n_{\gamma_3 \gamma_4} \; ,
\end{equation}
where $V^{\rm ee}$ is an effective on-site Coulomb interaction
expressed in terms of Slater integrals that are linked to the
intra-atomic repulsion $U$ and exchange $J$, see
Eq.~(3) in Ref.~\onlinecite{shick99}. The essential feature of the generalized total
energy functional~(\ref{eq:eeHam}) is that it contains
spin-off-diagonal elements of the on-site occupation matrix
$n_{\gamma_1 \gamma_2} \equiv n_{m_1 \sigma_1,m_2 \sigma_2}$
which become important in the presence of large SOC.

For a given set of spin-orbitals $\{ \phi_{m \sigma} \}$, we
minimize the DFT+U total energy functional. It gives the
Kohn--Sham equations for a
two-component spinor $\bm{\Phi}_{i} = \left( \begin{array}{c} \Phi^{\uparrow}_i \\ \Phi^{\downarrow}_i \end{array} \right)$,
                                 
\begin{equation}
\sum_{\beta} \Bigl[ -\nabla^{2} + \hat{V}_{\rm eff} +  \xi\, ( \bm{l}
\cdot\bm{s} )  \Bigr]_{\alpha,\beta}
\Phi^{\beta}_i(\bm{r}) = e_{i} \Phi^{\alpha}_{i}(\bm{r}) \; ,
\label{eq:KohnSham}
\end{equation}
where the effective potential $\hat V_{\rm eff}$ is a sum of the standard
(spin-diagonal) DFT potential and the on-site electron-electron
interaction potential $V_{U}$,

\begin{equation}
\label{eq:VU} \hat{V}^{\alpha,\beta}_{U} = \sum_{m, m'}
  |\phi_{\alpha m} \rangle \Bigl(W^{\alpha m, \beta m'} - \delta_{m, m'}
  \delta_{\beta,\alpha} W_{\rm dc}^{\alpha}\Bigr) \langle \phi_{\beta m'}|\,,
\end{equation}
where
\begin{eqnarray}
W^{\alpha m, \beta m'} =  \sum_{p \sigma,q \sigma'}
 \Big(  \langle m' \beta, p \sigma |V^{ee}|m \alpha ,q \sigma' \rangle \Big. \nonumber \\
 \Big. -  \langle m' \beta ,p \sigma|V^{ee}|q \sigma',m \alpha \rangle
\Big) n_{p \sigma,q \sigma'}
\end{eqnarray}
and $W_{\rm dc}^{\alpha}$ is the
double-counting correction. 
The most commonly used form of $W^{\sigma}_{\rm dc}$ is 
the so-called
``fully localized'' (or atomic-like) limit (FLL)~\cite{LAZ1995},
$W^{\sigma}_{\rm dc}= U(n_f - 1/2)
- J(n^{\sigma}_f  - 1/2)$.
Another form of the DFT+U functional is often called as ``around-mean-field'' (AMF) limit of the DFT+U~\cite{AZA1991},
$W^{\sigma}_{\rm dc}= U n^{-\sigma}_f + \frac{2l}{(2l+1)}(U-J) n^{\sigma}_f$. 
The operator $| \phi_{\alpha m} \rangle \langle \phi_{\beta m'}|$
in Eq.~(\ref{eq:VU}) acts on the two-component spinor wave function
$\bm{\Phi}$ as 
$|\phi_{\alpha m} \rangle \langle \phi_{\beta m'}|\Phi^{\beta}
\rangle$.

In addition to the spin-dependent DFT potential, the DFT+U method
creates a spin- and orbitally-dependent on-site ``$+U$'' potential,
which enhances orbital polarization beyond the polarization given by
the DFT alone (where it comes from the SOC only). We also note that the DFT
contributions to the effective potential 
$\hat{V}_{\rm eff}$ in Eq.~(\ref{eq:KohnSham}) are corrected to exclude
the double-counting of the $f$-states nonspherical contributions to
the DFT and DFT+U parts of the potential.
The nonspherical part of the DFT potential is expanded 
in terms of the lattice harmonics~$K_{\nu}$,
$V^{\rm NSH}_{\rm DFT}(\bm{r}) = \sum_{\nu} V_{\nu}(r)
K_{\nu}(\hat{\bm r})$. The DFT contributions to 
the muffin-tin nonspherical matrix elements, that are proportional to $\langle
lm_1|K_{\mu}|l m_2 \rangle$ for $l=3$ orbital quantum number, are
removed.

\subsection{DFT combined with the Anderson impurity model (DFT+ED)}
\label{sec:LDA+ED}

To proceed beyond DFT+U in the electronic structure of the
4$f$ adatoms on graphene, we
make use of the ``DFT++'' methodology~\cite{A.I.Lichtenstein1998}. We consider 
the  one-particle Hamiltonian found from \textit{ab initio} electronic
structure calculations plus  the
on-site Coulomb interaction  describing the $f$-electron correlation
of an adatom. The effects of the Coulomb interaction 
on the electronic structure are described by a one-particle
selfenergy $\Sigma(z)$ (where $z$ is a (complex) energy),
which is calculated in a multiorbital Anderson impurity model, 
\begin{eqnarray}
\label{eq:hamilt}
H_{\rm imp}  =  \sum_{\substack{k m m'  \\ \sigma \sigma'}}
 [\epsilon^{k}]_{m m'}^{\sigma \; \; \sigma'} b^{\dagger}_{km\sigma}b_{km'\sigma'}
 +\sum_{m\sigma} \epsilon_f f^{\dagger}_{m \sigma}f_{m \sigma}
\nonumber \\
 + \sum_{mm'\sigma\sigma'} \bigl[\xi\,(\bm{l}\cdot\bm{s})
  + \Delta_{\rm CF}\bigr]_{m m'}^{\sigma \; \; \sigma'}
  f_{m \sigma}^{\dagger}f_{m' \sigma'}
\nonumber \\
 +  \sum_{\substack{k m m'\\ \sigma \sigma'}}   \Bigl(
[V^{k}]_{m m'}^{\sigma \; \; \sigma'}
 f^{\dagger}_{m \sigma} b_{k m'  \sigma'} + {h.c.}
  \Bigr)
\nonumber \\
 + \frac{1}{2} \sum_{\substack{m m' m'' \\ m''' \sigma \sigma'}}
  U_{m m' m'' m'''} f^{\dagger}_{m\sigma} f^{\dagger}_{m' \sigma'}
  f_{m'''\sigma'} f_{m'' \sigma}.
\end{eqnarray}
Here $f^{\dagger}_{m \sigma}$ creates an electron in the 4$f$ shell and
$b^{\dagger}_{m\sigma}$ creates an electron in the ``bath''
that consists of those host-band states that hybridize with the
impurity 4$f$ shell.
The energy position $\epsilon_f$ of the impurity level, and the bath
energies $\epsilon^{k}$ are measured from the chemical potential $\mu$.
The parameters $\xi$ and $\Delta_{\rm CF}$ specify
the strength of the SOC and the size of the crystal field at the impurity.
The parameter matrices  $V^{k}$ describe the hybridization between the
$f$ states and the bath orbitals at energy $\epsilon^{k}$.

The band Lanczos method~\cite{J.Kolorenc2012}, paired with an efficient truncation of the many-body Hilbert space~\cite{Kolorenc2015} is employed to
find the lowest-lying eigenstates of the many-body Hamiltonian
$H_{\rm imp}$  for a given number $n_f$ of correlated electrons, and to calculate 
the one-particle Green's function $[G_{\rm imp}(z)]_{m m'}^{\sigma \; \; \sigma'}$
in the subspace of the $f$ orbitals. 
The selfenergy $[\Sigma (z)]_{m m'}^{\sigma \; \; \sigma'}$ is then
obtained from the inverse of the Green's function matrix
$G_{\rm imp}$.

Once the selfenergy is known, the local Green's function $G(z)$ for
the electrons in the $f$ manifold of the rare-earth adatom is
calculated as
\begin{equation}
\label{eq:gf}
G(z) =  \Big[ {G}_{0}^{-1}(z) + \Delta \epsilon - \Sigma(z) \Big]^{-1} \, , 
\end{equation}
where  ${G}_{0}(z)$ is the noninteracting Green's function, and
$\Delta \epsilon$ is chosen to ensure 
that $n_f = \pi^{-1} \Im \Tr\int_{-\infty}^{E_{{\rm F}}} {\rm d} E\,
G(E-{\rm i} 0)$ is equal to the given number of correlated electrons.
Subsequently, we evaluate the occupation matrix in the 4$f$ shell,
$n_{\gamma_1 \gamma_2} = \pi^{-1}\Im
\int_{-\infty}^{E_{\rm{F}}} {\rm d} E \, [G(E-\rm i0)]_{\gamma_1
  \gamma_2}$.
This matrix is used to construct an effective DFT+U
potential ${V}_{U}$, Eq.~(\ref{eq:VU}), which is inserted into
Kohn--Sham Eqs.~(\ref{eq:KohnSham}). The DFT+U Green's
function $G_{U}(z)$ is evaluated from the eigenvalues and
eigenfunctions of Eq.~(\ref{eq:KohnSham}), represented in the FLAPW
basis, and then it is used to calculate an updated noninteracting Green's function
${G}_{0}^{-1}(z)= G_{U}^{-1}(z) + V_{U}(z)$.
In each iteration, a new value of the 4$f$-shell
occupation is obtained. Subsequently,
a new self-energy $\Sigma(z)$ corresponding to the updated $f$-shell
occupation is 
constructed. Finally, the next iteration is started by evaluating the new
local Green's function, Eq.~(\ref{eq:gf}). The steps are iterated
until self-consistency over the charge density is reached.

When the hybridization between the $f$ states and the bath orbitals 
is weak, one can neglect the first and fourth terms in  Eq.~(\ref{eq:hamilt}),
and the Anderson impurity model is reduced to the atomic model. This approximation 
is called Hubbard-I approximation (HIA). The use of HIA allows us to
substantially reduce the computational
cost needed for the exact diagonalization of the Hamiltonian~(\ref{eq:hamilt}). 
The same procedure for the charge-density self-consistency is used for DFT+HIA.
Further details of the DFT+HIA implementation  in the FP-LAPW basis are described in Ref.~\onlinecite{shick09}. 

\section{Samarium on graphene}
\label{sec:Sm}

\subsection{DFT+U}

We start with the application of the DFT+U approach to Sm@GR. The Slater
integrals that define the on-site Coulomb interaction are chosen as
$F_0=6.87$~eV, $F_2=9.06$~eV, $F_4=6.05$~eV, and $F_6=4.48$~eV.  They
correspond to Coulomb~$U=6.87$~eV and Hund exchange~$J=0.76$~eV. The
spin ($M_S$) and orbital ($M_L$) magnetic moments are given in
Table~\ref{tab:1} together with the occupation of the Sm $4f$ orbitals
$n_f$. In these calculations, the magnetization (spin${}+{}$orbital)
is constrained along the crystallographic axes: $x, y$ (in plane), and
$z$ (out of plane). The DFT+U-FLL yields a solution with both
$M_S$ and $M_L$ non-zero, and $n_f$ very close to six.
Thus the FLL flavor of the DFT+U gives an $f^6$ magnetic ground state
with the  total moment $M_J$ = 2.9  $\mu_B$.
On the contrary, the DFT+U-AMF converges to a practically nonmagnetic
$f^6$ ground state with all $M_S$, $M_L$ and $M_J$ close to zero
(Table~\ref{tab:1}).

\begin{table} 
\caption{Spin ($M_S$), and orbital ($M_L$) magnetic moments (in $\mu_B$) and $4f$ occupation $n_f$ of
  the Sm adatom on graphene for three different directions of the
  magnetization $M$: $x, y$ (in plane), and $z$ (out of plane).}
\begin{center}

\label{tab:1}

\begin{ruledtabular}
\begin{tabular}{cccc|ccc}

 \multicolumn{1}{c}{{\bfseries Sm@GR}}&\multicolumn{3}{c}{FLL} &\multicolumn{3}{c}{AMF} \\
\hline
        & $n_f$ & $M_S$& $M_L$ & $n_f$ & $M_S$& $M_L$\\
$M || x$     & 5.94    & 5.85     & $-2.90$    &  5.94  & 0.09    & $-0.04$ \\
$M || y$     & 5.94    &5.86      & $-2.91$    & 5.94   & 0.09    &$-0.03$ \\
$M || z$     & 5.94  &5.84      &$-2.93$     & 5.94   & 0.19 &$-0.10$\\

\end{tabular}
\end{ruledtabular}

\end{center}
\end{table}

\begin{figure}[htbp]
\centerline{\includegraphics[width=1.0\columnwidth]{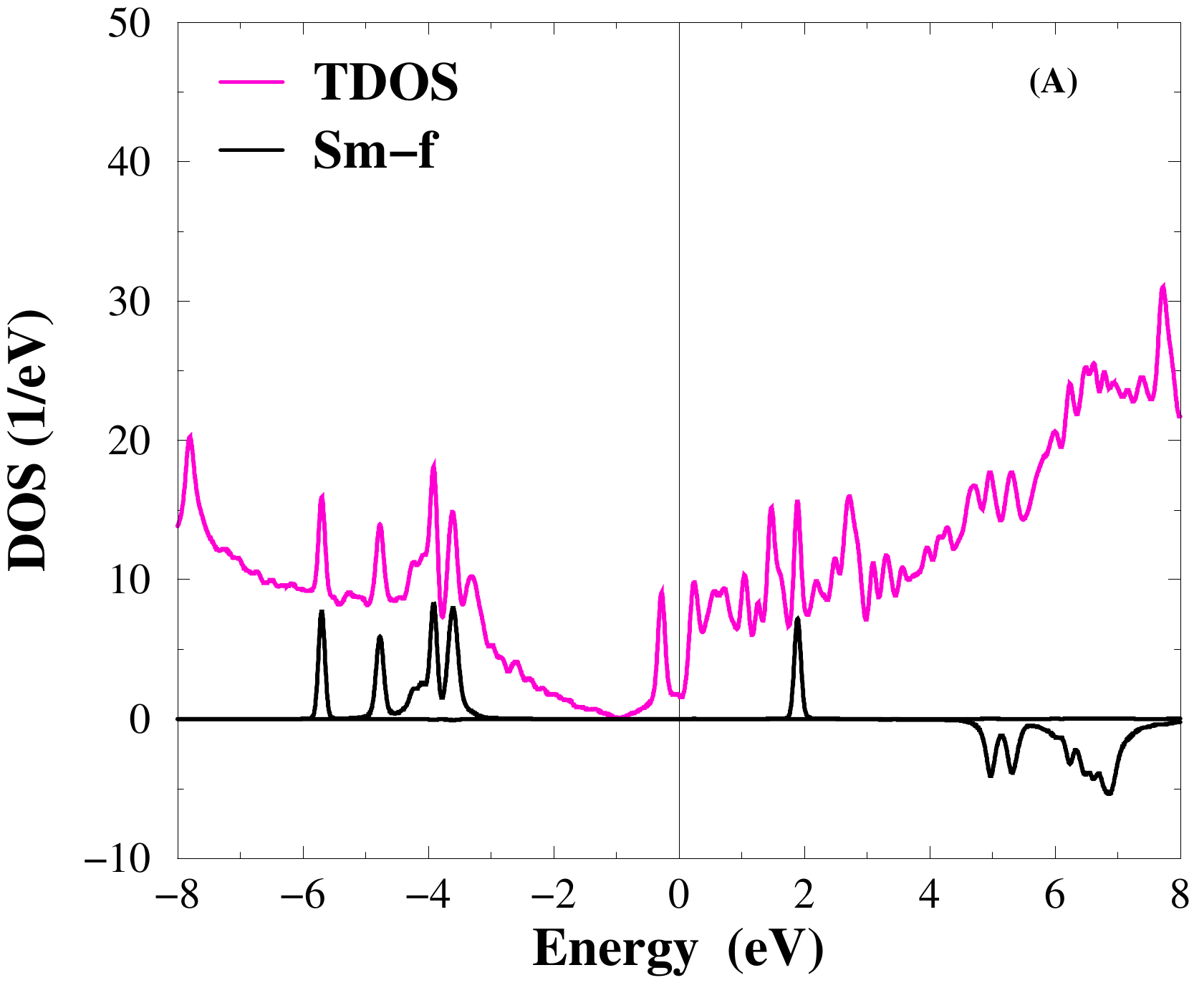} }
\centerline{\includegraphics[width=1.0\columnwidth]{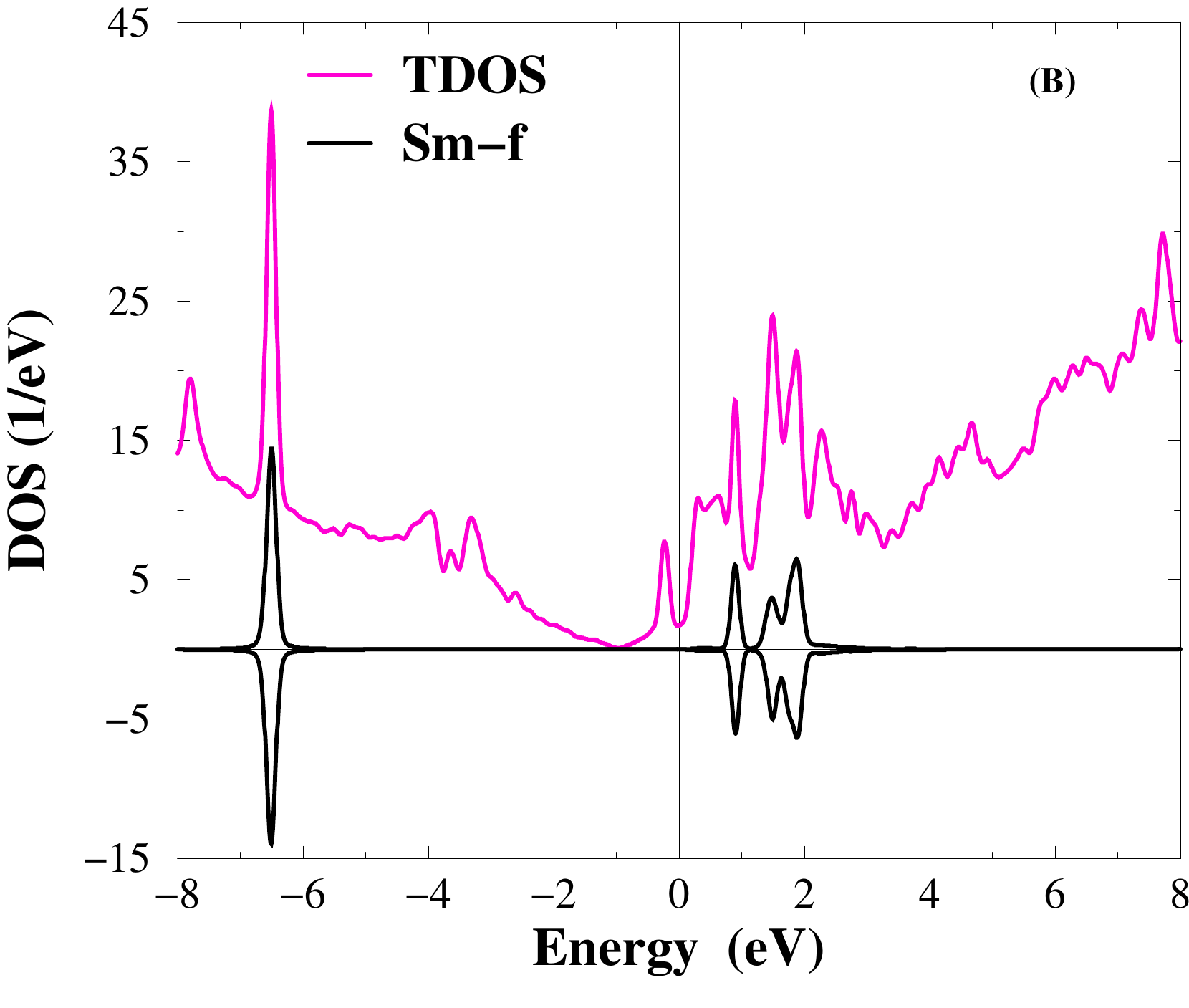} }
\caption{The total (TDOS) and spin-resolved $f$-orbital
  density of states for the Sm adatom on graphene calculated with
  DFT+U-FLL (A) and DFT+U-AMF (B).}
\label{fig:2}
\end{figure}

The calculated total density of states (TDOS, for both spins, and per
unit cell) and the $f$-orbital spin-resolved DOS for Sm adatom calculated
with DFT+U-FLL and DFT+U-AMF are shown in Fig.~\ref{fig:2}.  The
DFT+U-FLL yields a mean-field solution with broken symmetry. This is
because the part of the Coulomb interaction treated in the
Hartree--Fock-like approximation is transformed into the exchange
splitting field. This exchange field is several eV strong (see
Fig.~\ref{fig:2}) and by far exceeds any imaginable external
magnetic field. This exchange field is reduced to almost zero in the
DFT+U-AMF calculations. The DFT+U is not based on any kind of
atomic coupling scheme ($LS$ or $jj$), since it determines a set of
single-particle orbitals that variationally minimize the total energy.
The AMF calculated $f^6$ nonmagnetic ground state corresponds to the
Slater determinant formed of six equally populated $j=5/2$ orbitals.

\subsection{DFT+HIA}

The observation that two different flavors of DFT+U yield different
results for the magnetic properties is similar to $fcc$-Am where the
DFT+U results  strongly depend on the choice of the DFT+U double
counting~\cite{shick2006}. This situation is quite alarming, and
indicates that one has to go beyond the static mean-field approximation
to accurately model these systems. Such an improved
approximation was introduced in Sec.~\ref{sec:LDA+ED} where the
Coulomb potential $V_U$, Eq.~(\ref{eq:VU}), is calculated from the
occupation matrix~$n_{m_1 \sigma_1,m_2 \sigma_2}$ corresponding to a
multi-reference many-body wave function instead of a single Kohn--Sham
determinant. The many-body wave function is the ground state of the
impurity model from Eq.~(\ref{eq:hamilt}) with the following parameters:
the Slater integrals are the same as those used in the DFT+U calculations,
the spin-orbit parameter $\xi=0.16$~eV was determined from DFT
calculations, and the crystal-field effects are neglected,
$\Delta_{\rm CF}=0$.

First, we excluded the hybridization between the $f$ states and the bath orbitals
in Eq.~({\ref{eq:hamilt}), and used DFT+HIA.
The occupation of the 4$f$~shell self-consistently determined from
Eq.~({\ref{eq:KohnSham}) is $\langle n_f \rangle=5.95$ (FLL double
counting) and $\langle n_f \rangle=5.98$ (AMF double
counting). When we alternatively fix $\epsilon_f$ in Eq.~({\ref{eq:hamilt}) to $-W_{\rm dc}$ from Eq.~(\ref{eq:VU}),  we obtain the occupation
$\langle n_f \rangle=6.0$. It means that all $f$-electrons of Sm are fully localized.
The ground state of  the 4$f$~shell is  a nonmagnetic singlet with all angular
moments equal to zero ($S = L = J =0$). The $f$-orbital DOS obtained from Eq.~(\ref{eq:gf}) is shown in
Fig.~\ref{hybridization}~(A). There is practically no
difference between the different double-counting variants,
FLL or AMF, in Eq.~(\ref{eq:KohnSham}).

\subsection{DFT+ED}

Next, we determine the bath parameters $V^{k}$ and $\epsilon^{k}$,
assuming that the DFT 
represents the noninteracting model. That is, we associate the DFT Green's
function ${G}_{\rm{DFT}}(z)$ with the
Hamiltonian~(\ref{eq:hamilt}) when the coefficients of the Coulomb
interaction matrix are set to zero ($U_{mm'm''m'''}=0$). The
hybridization function ${\Delta(\epsilon)}$ is then estimated as
${\Delta}(\epsilon) =
 \Im \mathop{\rm Tr} [G^{-1}_{\rm DFT}(\epsilon - {i0})]$.

A detailed inspection shows
that the hybridization matrix is, to a good approximation, diagonal in the
$\{j,j_z\}$ representation. Thus, we assume the first and fourth terms
in Eq.~(\ref{eq:hamilt}) to be diagonal in
$\{j,j_z\}$. Hence we only need to specify one bath state (six
orbitals) with $\epsilon^{k=1}_{j=5/2}$ and $V^{k=1}_{j=5/2}$, and
another bath state (eight orbitals) with $\epsilon^{k=1}_{j=7/2}$
and $V^{k=1}_{j=7/2}$. Assuming that the most important
hybridization occurs in the vicinity of the Fermi level $E_{\rm F}$, the numerical values
of the hybridization parameters $V^{k=1}_{5/2,7/2}$
are found from the relation~\cite{Gunnarsson89}
$\pi \sum_{k} {|V_{j}^{k}|}^2 \delta(\epsilon_{j}^{k} - \epsilon) = 
- \Delta(\epsilon)/N_j$ averaged over the energy interval,
$E_{\rm F} - 0.5$ eV $\le \epsilon \le E_{\rm F} + 0.5$ eV,
with $N_j=6$ for $j=5/2$ and $N_j=8$ for $j=7/2$.
The bath-state energies
$\epsilon^{k=1}_{5/2,7/2}$ shown in Table~\ref{parameters} are then
adjusted to approximately reproduce
the DFT occupations of the $f$ states, $n_f^{5/2}$ and $n_f^{7/2}$.
Note that the magnitudes of the hybridization parameters $V$ are very
small indicating the localized nature of the 4$f$-states.   

\begin{table} 

\caption{$f$-states occupations  $n_f^{5/2}$ and $n_f^{7/2}$, and
  bath-state parameters $\epsilon^1_{5/2}$,
  $\epsilon^1_{7/2}$, $V^{1}_{5/2}$, $V^{1}_{7/2}$ (all energies in
  eV) for Sm and Nd adatoms determined from DFT calculations.} 
\begin{center}
\begin{ruledtabular}
\begin{tabular}{lcccccc}

Adatom & $n_f^{5/2}$ &$n_f^{7/2}$&$\epsilon^1_{5/2}$&$V^{1}_{5/2}$&$\epsilon^1_{7/2}$&$V^{1}_{7/2}$   \\
\hline
Sm   & 5.72 & 0.36  &0.025 &0.071  & $-0.500$   & 0.077 \\
Nd    & 3.49 & 0.14 &0.050  &0.085  &$-0.500$    & 0.087 \\

\end{tabular}
\end{ruledtabular}
\end{center}
\label{parameters}
\end{table}

The occupation of the 4$f$~shell self-consistently determined from
Eq.~({\ref{eq:KohnSham}) is $\langle n_f \rangle=5.95$ (FLL double
counting) and $\langle n_f \rangle=5.97$ (AMF double
counting). Since the occupation 
is very close to $\langle n_f \rangle=6.0$, we kept $\epsilon_f$ in Eq.~({\ref{eq:hamilt}) at $\langle
  n_f \rangle=6.0$. The ground state of the cluster formed by the
4$f$~shell and the bath is a nonmagnetic singlet with all angular
moments equal to zero ($S = L = J =0$). In this ground state, there
are $\langle n_f \rangle=6.0$ electrons in the $4f$ shell and $\langle
n_{\rm bath} \rangle=8.0$ electrons in the bath states.
The ground-state expectation values of the angular moments of the $4f$
shell are calculated as $S_f = 2.92$, $L_f = 2.92$, and
$J_f = 0.03$.  The singlet ground state is separated from the first
excited state (triplet) by a gap of 50~meV. The $f$-orbital density
of states obtained from Eq.~(\ref{eq:gf}) is shown in
Fig.~\ref{hybridization}~(B). Comparison with DFT+HIA,
Fig.~\ref{hybridization}~(A), demonstrates similar features
with about 1 eV upward energy shift.  Also, we have examined 
the double-counting choice in Eq.~(\ref{eq:KohnSham}), and found
practically no difference between the different double-counting variants,
FLL or AMF.

\begin{figure}[tbp]
\centerline{\includegraphics[angle=0,width=1.0\columnwidth]{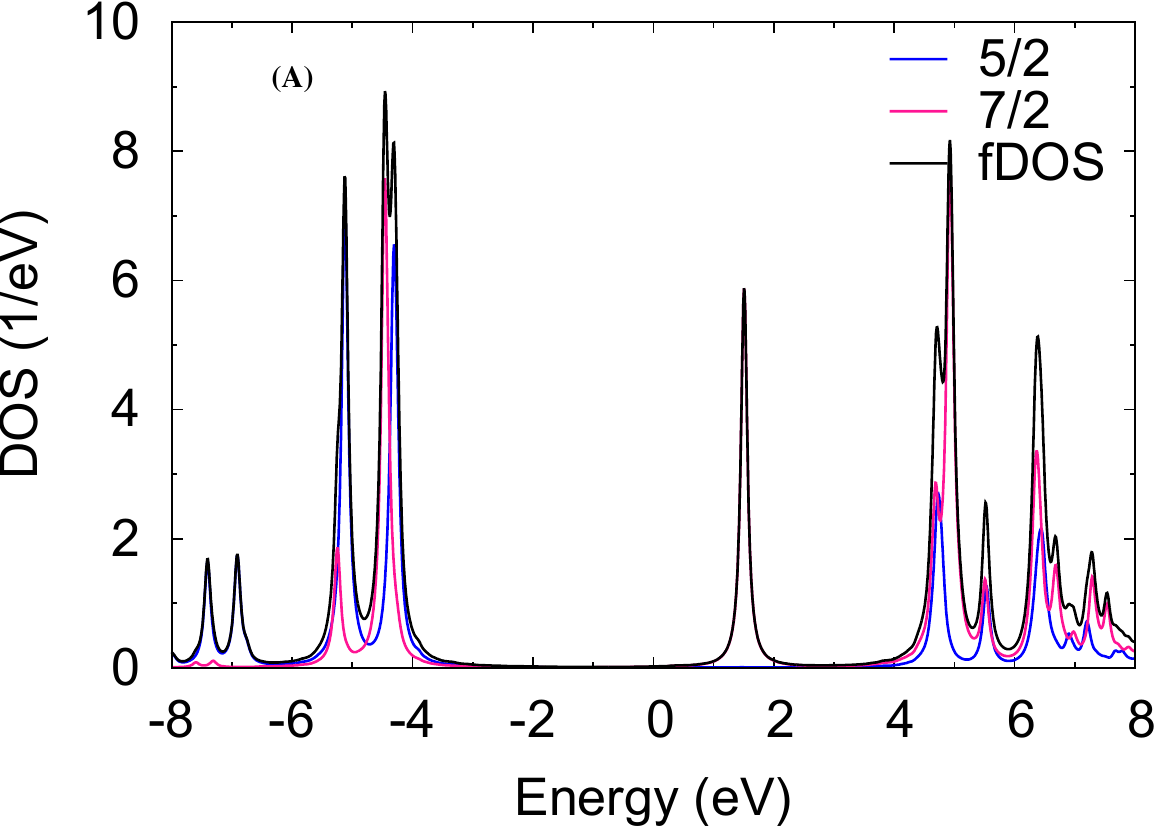}}
\centerline{\includegraphics[angle=0,width=1.0\columnwidth]{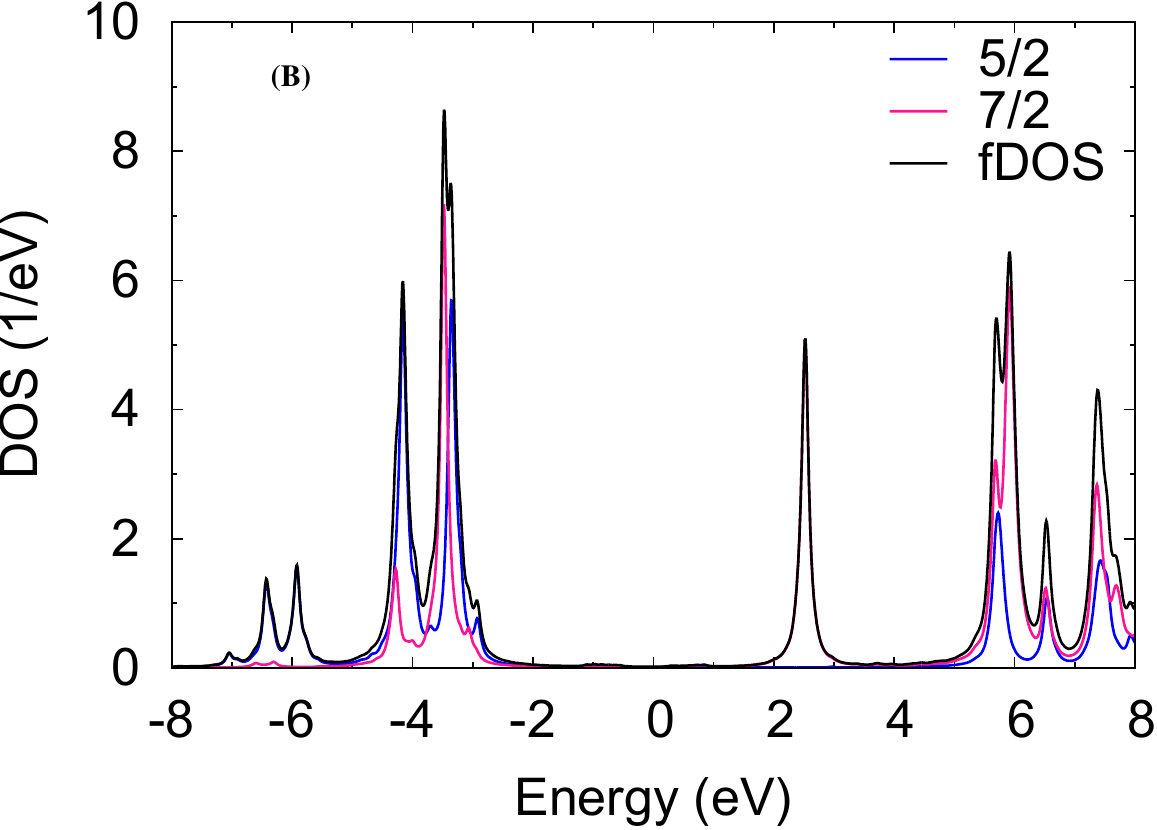}}
\caption{$f$-electron density of states (fDOS, and $j = 5/2$, $7/2$
projected) for the Sm atom in Sm@GR resulting from DFT+HIA calculations (A); 
fDOS, and $j = 5/2$, $7/2$
projected fDOS, for the Sm atom in Sm@GR from DFT+ED~(B).}
\label{hybridization}
\end{figure}

\subsection{Implications for x-ray absorption spectroscopy}

Information about the $4f$ states can be gleaned from the x-ray absorption 
spectroscopy (XAS). In these experiments, the intensities $I_{5/2}$
($3d_{5/2} \rightarrow 4f_{5/2,7/2}$) and $I_{3/2}$ ($3d_{3/2}
\rightarrow 4f_{5/2}$) of the individual absorption lines are measured
and the branching ratio $B = I_{5/2}/(I_{3/2} + I_{5/2})$ is
evaluated~\cite{KMoore2009}. We compute the branching ratio $B$ for
core to valence $3d$--$4f$ transition by obtaining $n_{5/2}$ and
$n_{7/2}$ from the local occupation matrix $n_{\gamma_1 \gamma_2}$ and
making use of the sum rule~\cite{KMoore2009},
\begin{equation}
B = \frac{3}{5} \; - \; \frac{2}{5}\, \frac{n_f^{7/2} -\frac{4}{3}\, n_f^{5/2}}{14 - n_f} \, .
\label{eq:BR}
\end{equation}

\begin{table}[t] 
\caption{Occupation $n_f$,   values of $n_f^{5/2}$ and $n_f^{7/2}$, and branching ratio $B$  for Sm@GR.
The atomic theory values~\cite{KMoore2009} for  $n_f=6$ in the $LS$ and $jj$ coupling schemes are also given.} \label{tab:occupations}
\centering

\begin{ruledtabular}
\begin{tabular}{lcccc}

{\bf Sm@GR}  & $n_f$  & $n_f^{5/2}$ &  $n_f^{7/2}$& $B$  \\
\hline
DFT+U-FLL    & 5.94 & 3.33 & 2.60 & 0.69 \\
DFT+U-AMF  & 5.94 & 5.87 & 0.07 & 0.985 \\
DFT+HIA-FLL & 5.95 & 3.80 & 2.14 & 0.745 \\
DFT+ED-FLL  & 5.95 & 3.81 & 2.14 & 0.75\\
DFT+HIA-AMF & 5.98 & 3.82 &2.16 & 0.75\\
DFT+ED-AMF &5.97 & 3.82 & 2.16 & 0.75\\
atomic $LS$       &  6   & 3.14 &2.86  & 0.67\\
atomic $jj$        &  6   &6.00  & 0.00 & 1.00\\

\end{tabular}
\end{ruledtabular}
\end{table}

The DFT+U-FLL as well as DFT+HIA and DFT+ED yield
the branching ratios close to the atomic $LS$-coupling limit
(Table~\ref{tab:occupations}). On the contrary, the DFT+U-AMF value is
close to the $jj$-coupling atomic value $B=1.0$. It is rather well established that the rare-earth atoms with the localized $f$ shell are well
described by the $LS$-coupling scheme, and our DFT+HIA and DFT+ED calculations, which are not bound by any particular
atomic coupling scheme, illustrate once again the validity of the
conventional atomic theory. At the same time, the DFT+U-AMF does not
have the proper atomic limit since it is very far from the
$LS$-coupling scheme.

As we have shown, the use of  DFT+U for Sm on graphene can lead to erroneous conclusions about the magnetic character of the Sm adatom.
In fact, recent DFT+U calculations~\cite{YaJingLi2016} for the rare-earth atoms embedded in graphene, including Sm, report it to carry
large spin and orbital magnetic moments. We think that the magnetic character of Sm atom in graphene was not determined correctly~\cite{YaJingLi2016}.

\section{Neodymium on graphene}
\label{sec:Nd}

\subsection{DFT+HIA}

Theoretical evaluation of the local magnetic moments of the rare-earth atoms  adsorbed on a nonmagnetic substrate is an important issue
in the context of creating a single 4$f$-atom
magnet~\cite{Miyamachi2013, Donati2014}. As an example of the
rare-earth adatom,
where the local moment is expected to exist from the atomic $LS$-coupling scheme arguments,      
we consider the case of Nd@GR. For the DFT+HIA calculations, the Slater
integrals  $F_0=6.76$~eV, $F_2=9.06$~eV, $F_4=6.05$~eV, and $F_6= 4.48$~eV were chosen. They corresponds to 
Coulomb~$U=6.76$~eV and exchange~$J = 0.76$~eV. The spin-orbit
parameter was determined by DFT that yields $\xi=0.13$~eV.  
The bath parameters  were evaluated using the same procedure as for Sm@GR, they are listed in  Table~\ref{parameters}.
It is seen that the  hybridization strength in Nd@GR is rather similar
to Sm@GR. This weak hybridization allows us to use the simpler
DFT+HIA method.

The ground state of the Nd atom on graphene, the solution
of Eq.~({\ref{eq:KohnSham}), has  $\langle n_f \rangle=3.66$
$f$ electrons.
Note that Nd atom in solid-state compounds commonly has a valency $3+$, and the
deviation from the atomic-like $f^4$ configuration is thus not surprising.
The ground state has degeneracy of nine, and the expectation values of
the $4f$-shell moments are $S_f = 1.96$, $L_f = 5.95$, and $J_f = 4.00$.
These values are consistent with  the   $^5I_4$  $LS$-coupled  $f^4$ atomic ground state.
The degenerate character of the ground state dictates the presence of
local moment for Nd@GR.  The XAS branching ratio
$B=0.715$ is calculated, and can be verified experimentally.

\begin{table}[t] 
\caption{Spin ($M_S$), and orbital ($M_L$) magnetic moments (in $\mu_B$) and $4f$ occupation $n_f$ of
 Nd@GR for three different directions of the magnetization $M$: $x, y$
 (in plane) and $z$ (out of plane).}
\label{tab:Nd}
\begin{center}
\begin{ruledtabular}
\begin{tabular}{cccc}
{\bfseries Nd@GR}

                                 & $n_f$ & $M_S$& $M_L$ \\
\hline
$M || x$                    & 3.78    &  3.70    & $-4.59$   \\                                        
$M || y$                    & 3.78    &  3.71    & $-4.60$   \\
$M || z$                    & 3.78    &  3.69    & $-2.58$   \\
\end{tabular}
\end{ruledtabular}
\label{table2}
\end{center}

\end{table}

\subsection{DFT+U}

Since we have shown that DFT+U+AMF does not have the correct atomic
limit (it is not close to the $LS$-coupling scheme),  
we apply only the  DFT+U-FLL  approach  to Nd@GR. It yields  non-zero 
spin $M_S$ and orbital $M_L$ magnetic moments, which are given in Table~\ref{table2}  together with
the occupation of the Nd adatom $f$ orbitals, $n_f$. In these
calculations, the magnetization (spin${}+{}$orbital)
is constrained along the crystallographic axes:  $x, y$ (in plane) and
$z$ (out of plane). 
Note that $M_S$ and $M_L$ have different physical meaning than $S_f$ and $L_f$ in DFT+ED calculations:
they represent the projections of the spin and orbital moments on the selected axis, while $S_f$, and $L_f$ are 
the expectation values of the many-body spin and orbital operators squared.
Qualitatively, one can say that these DFT+U solutions represent
different mean-field approximations
(or their linear combinations) of the degenerate many-body ground state.  

The $f$-orbital DOS obtained in DFT+HIA
calculations is shown in Fig.~\ref{fig:dos}(A). Comparison with
DFT+U, see Fig.~\ref{fig:dos}(B), shows that DFT+U gives rather correct 
placement of the $f$-states. No multiplet splittings, which are clearly seen in
the DFT+HIA DOS are resolved in DFT+U. This is expected from
the single-determinant DFT+U approximation. Both DFT+HIA and
DFT+U suggest no $f$-character  DOS  in the vicinity of $E_{\rm F}$.

\begin{figure}[tbp]
\centerline{\includegraphics[angle=0,width=1.0\columnwidth]{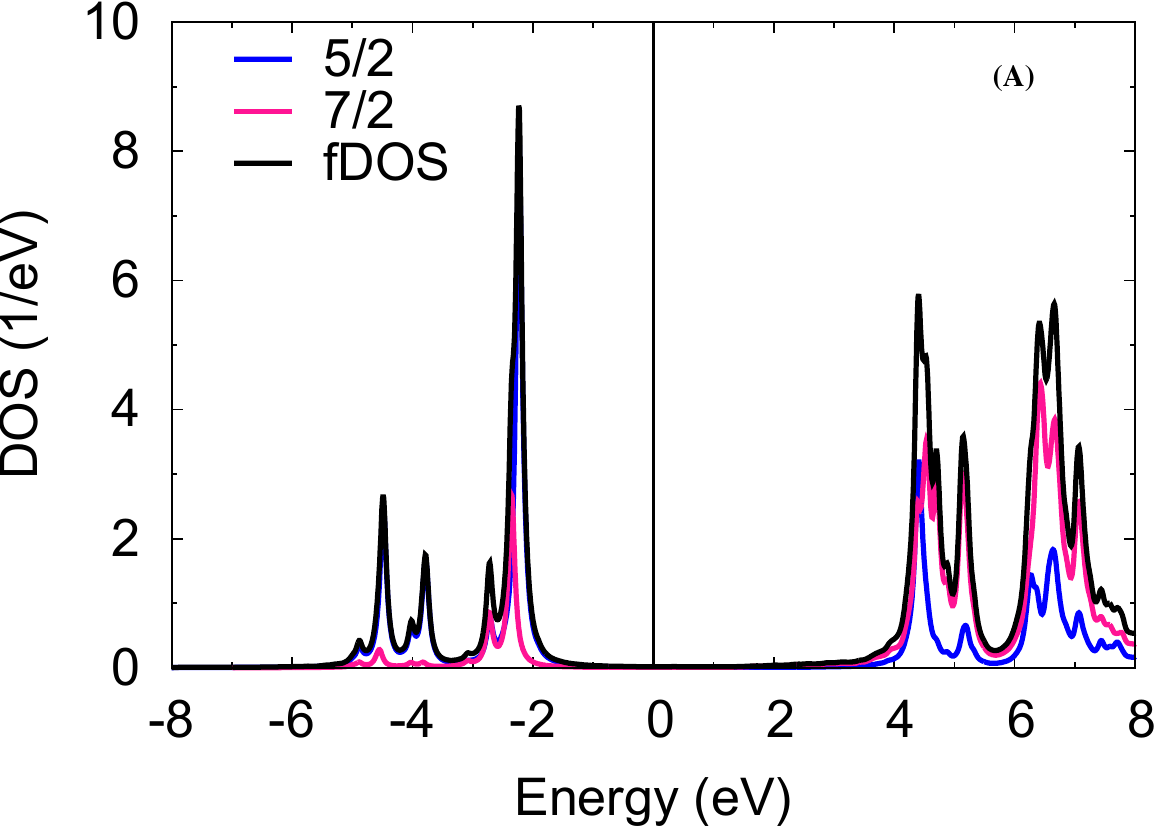}}
\centerline{\includegraphics[angle=0,width=1.0\columnwidth]{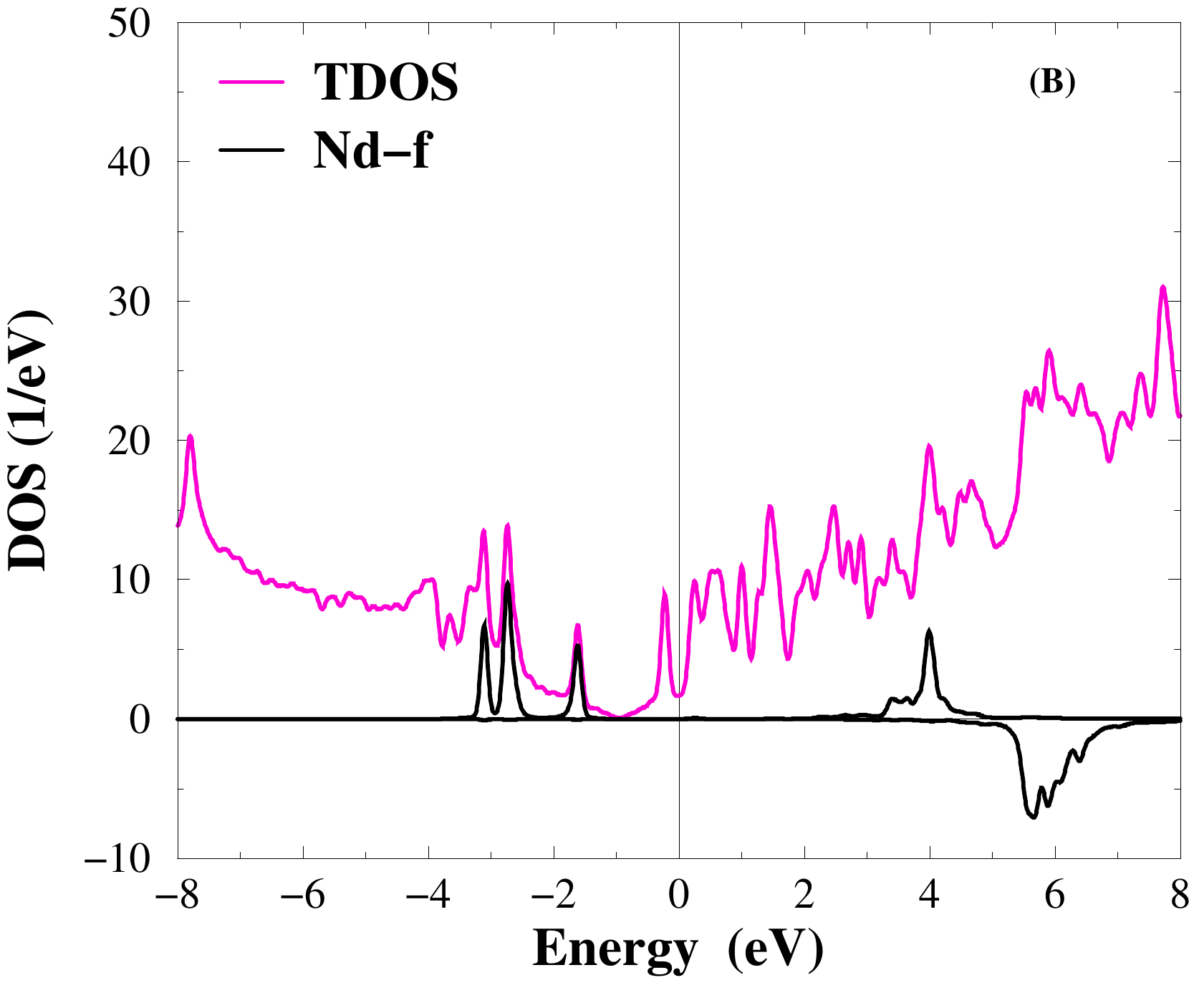}}
\caption{$f$-electron density of states (fDOS, and $j$ = 5/2, 7/2
projected) for the Nd atom in Nd@GR calculated with DFT+HIA (A);
the total (TDOS) and spin-resolved $f$-orbital DOS for the  Nd adatom
in Nd@GR calculated with DFT+U-FLL (B).}
\label{fig:dos}
\end{figure}

\section{Conclusions}

The electronic structure and magnetic properties of Sm and Nd impurities on a
free-standing graphene were investigated making use of DFT+U, DFT+HIA and DFT+ED
methods in order to analyze the role of the electron correlations and
the spin-orbit coupling. DFT+U calculations result in non-zero local magnetic moments for both adatoms.  This is expected for Nd, but not for
Sm, which has a nonmagnetic $f^6$  ($J=0$) ground state
configuration.  Application of the DFT+HIA and DFT+ED methods solves this
problem, and yields a nonmagnetic singlet ground state with
$n_f=6.0$, and $J=0$ for the Sm adatom, while the degenerate ground
state of  Nd adatom retains the local magnetic moment with  $n_f=3.7$,
and $J=4.0$. Our results show that the DFT+U predictions for the $f$ systems 
close to the atomic limit should be treated with caution, keeping in mind the 
ambiguities inherent to the DFT+U approximation. 

\section{Acknowledgments} 
We acknowledge stimulating discussions with P. {Jel\'\i nek}, M. Telychko  and L. Havela. 
Financial support was provided by the Czech Science Foundation (GACR) Grant No.~15-07172S, the National Science Centre (Poland) Grant No.~DEC-2015/17/N/ST3/03790,
the Deutsche Forschungsgemeinschaft (DFG) Grant No. DFG LI 1413/8-1.
Access to computing and storage facilities owned by parties and projects contributing to the National Grid Infrastructure MetaCentrum provided under the programme "Projects of Projects of Large Research, Development, and Innovations Infrastructures" (CESNET LM2015042), is appreciated.

\vspace{1cm}

\bibliography{refs}

\end{document}